\documentclass[%
 reprint,
 amsmath,amssymb,
 aps,
]{revtex4-1}

\usepackage{graphicx}
\usepackage{dcolumn}
\usepackage{bm}

\usepackage{amsmath}
\usepackage{color}
\usepackage{natbib}

\begin{document}


\title{Calculation of entanglement in graph states up to five-qubit
based on generalized concurrence
}

\author{Ahmad Akhound}
\email{aakhound@pnu.ac.ir}
\affiliation{%
Department of Physics, Payame Noor University, P.O. Box 19395-3697, Tehran, Iran.
}
\author{Saeed Haddadi}
\email{haddadi@physicist.net}
\affiliation{%
Department of Physics, Payame Noor University, P.O. Box 19395-3697, Tehran, Iran.
}
\author{Mohammad Ali Chaman Motlagh}
\email{machaman2000@yahoo.com}

\affiliation{%
Department of Physics, Payame Noor University, P.O. Box 19395-3697, Tehran, Iran.
}

\begin{abstract}
We propose a new classification for the entanglement in graph states based on generalized concurrence. The numerical results indicate that the eight different three-qubit graph states in three categories, 64 four-qubit graph states in five categories and 1024 five-qubit graph states are in ten classes. We also compare this classification with equivalence classes of these graph states under local complementation (LC) operator, and the obtained result suggests that classification by generalized concurrence is not in contradiction with the LC-rule.
\begin{description}

\item[PACS numbers]
03.65.Ud, 03.65.Mn, 03.67.-a.

\end{description}
\end{abstract}

\pacs{Valid PACS appear here}
\maketitle


\section{\label{sec:level1}INTRODUCTION}

Entanglement is a fundamental characteristic of quantum mechanics that reveals the important difference between classical and quantum physics. Entangled states indicate a variety of non-local quantum correlation subsystems \cite{Nielsen01}, which have many applications in quantum data, including quantum teleportation \cite{Bennett02,Furusawa03}, quantum dense coding \cite{Bennette04}, quantum cryptography and quantum computing \cite{Ekert05}. The entangled states have an essential role in qubit systems \cite{Akhound06} for usage in quantum information processing and communications as well \cite{Bose07,Christandle08}. The graph is one of the best mathematical tools to study some of the entangled states. The graph $ G=(V,E) $ contains a pair of limited set $ V \subseteq \mathbb{N} $ and $ E \subset V \times V $ \cite{Hein09,West10,Bondy11,Diestel12}. Members of the set $ V=\{{v_{1},\ldots,v_{n}}\} $ demonstrate the set of vertices and the set $ E=\{{e_{1},\ldots,e_{n}}\} $ which are edges or lines between the vertices \cite{Diestel12,Wu13}. It is broadly known that any graph state can be constructed on the foundation of a simple and undirected graph. We assign each vertex with a two-level quantum system(qubit), each edge represents the interaction between the corresponding two qubits and represent the Ising interaction between the qubits. Graph states are pure states which are used in quantum error correction, quantum computing models and quantum transport, the study of entanglement in qudit systems and investigation of non-local quantum \cite{Hein09,Eisert14}.

This paper is organized as follows: in section II we provide a relationship to determine the graph states of qubit-systems and then the generalized concurrence is introduced as a measurable quantity of entanglement in Sec. III. Sec. IV is dedicated to the classification of graph states under LC-rule and in the final section the results of these two categories will be discussed.

\section{\label{sec:level2}BASIC CONCEPTS}

The graph state corresponding to the graph $ G=(V,E) $   is obtained by different methods. In this paper, we use two ways. The first way to determination of $N$-qubit graph state is the following equation \cite{Chen15}

\begin{equation}
 |G\rangle=\frac{1}{\sqrt{2^{N}}}\sum_{\mu}(-1)^{\frac{1}{2} \mu \Gamma \mu^ {T}}|\mu\rangle,
\end{equation}

where $ \mu=(\mu_{1},\ldots,\mu_{N}) $ is a binary vector with $ \mu_{i}=0,1 $ for $ i=1,\ldots,N $ \cite{Qun16}, so that

\begin{equation}
 |G\rangle\in(\mathbb{C}^{2})^{\otimes V},
\end{equation}

where $\mathbb{C}^{2}$ represents a two-dimensional vector space with based vectors $ |0\rangle\doteq\binom{1}{0} $  and $ |1\rangle\doteq\binom{0}{1} $. The vectors are eigenstates of Pauli operator $ \sigma_{z} $. Pauli operators in this space are as follows

\begin{equation}
\sigma_{x}=\left(
             \begin{array}{cc}
               0 & 1 \\
               1 & 0 \\
             \end{array}
           \right),\,\sigma_{y}=\left(
             \begin{array}{cc}
               0 & -i \\
               i & 0 \\
             \end{array}
           \right),\,\sigma_{z}=\left(
             \begin{array}{cc}
               1 & 0 \\
               0 & -1 \\
             \end{array}
           \right).
\end{equation}

For a simple and undirected graph $ G=(V,E) $ with adjacency matrix is a square matrix $ \Gamma_{N\times N} $ so that it’s $ \Gamma_{ij}=\Gamma_{ji}=1 $ when there is an edge \{{$i$ , $j$\}} in the $E$, and $ \Gamma_{ij}=\Gamma_{ji}=0 $ when there is no edge \cite{Hein09,Diestel12}.

The second way to determination of $N$-qubit graph state, can be written as \cite{Hein09,Wu13,Eisert14}

\begin{equation}
 |G\rangle\:=\prod_{(i,j)\in E}CZ_{ij}|+\rangle_{x}^{\otimes N},
\end{equation}

where $ |+\rangle_{x}=\frac{1}{\sqrt{2}}(|0\rangle + |1\rangle ) $ is an eigenstate of Pauli operator $ \sigma_{x} $ with eigenvalue +1. Then for each edge connecting two qubits, $ i $ and $ j $, it is applied the $ CZ $ gate between qubits $ i $ and $ j $. This gate $ CZ_{ij} $ in Eq. (4) is as follows

\begin{equation}
CZ_{ij}\doteq\left(
               \begin{array}{cccc}
                 1 & 0 & 0 & 0 \\
                 0 & 1 & 0 & 0 \\
                 0 & 0 & 1 & 0 \\
                 0 & 0 & 0 & -1 \\
               \end{array}
             \right).
\end{equation}

In each $ N $-qubit system, the number of quantum states is numerous, whereas the number of simple graph state is $ 2^{\binom{N}{2}} $ \cite{Hein09}, so for a three-qubit system, there are 8 graph states, for four-qubit system there are 64 graph states and also for a five-qubit system there are 1024 graph states that their non-isomorphic graphs is plotted in figure 1. Two graphs $ G_{1}=(V_{1},E_{1}) $ and $ G_{2}=(V_{2},E_{2}) $ are called isomorphic $ (G_{1}\cong G_{2}) $ if there is a bijection $ f: V_{1}\rightarrow V_{2} $ is a mapping of a graph onto itself  between a set of vertices such that $ \{a,b\}\in E_{1} $ if and only if $ \{f(a),f(b)\}\in E_{2} $ \cite{West10,Bondy11,Diestel12}.

Using the Eq. (4), the $ N $-qubit graph state that corresponds to the empty graph is written as follows

\begin{equation}
|G_{0}\rangle\:=|+\rangle_{x}^{\otimes N},
\end{equation}

for example, the three-qubit graph state without edge that corresponds to the empty graph (graph No. 1 in figure 1) is defined as

\begin{align}
|G_{1}\rangle&=|+\rangle_{x}\otimes|+\rangle_{x}\otimes|+\rangle_{x}\nonumber\\
&=\frac{1}{\sqrt{8}}(|000\rangle+|001\rangle+|010\rangle+|011\rangle\nonumber\\
&\qquad+|100\rangle+|101\rangle+|110\rangle+|111\rangle).
\end{align}

In this method, by applying $ CZ $ gate between qubits $ 1 $ and $ 3 $ $(CZ_{1,3}) $ in Eq. (5) on state of Eq. (7), the three-qubit graph state that corresponds to the graph (No. 2) that there is an edge between vertices $ 1 $ and $ 3 $, it will be obtained

\begin{align}
|G_{2}\rangle&=\frac{1}{\sqrt{8}}(|000\rangle+|001\rangle+|010\rangle+|011\rangle\nonumber\\
&\qquad+|100\rangle-|101\rangle+|110\rangle-|111\rangle).
\end{align}

These results are also obtained with the first method. For this purpose, the adjacency matrix of the three-vertex graph with just a single edge between vertices 1 and 3 is written as follows

\begin{equation}
\Gamma_{1,3}=
 \left(
   \begin{array}{ccc}
     0 & 0 & 1 \\
     0 & 0 & 0 \\
     1 & 0 & 0 \\
   \end{array}
 \right).
\end{equation}

Now due to the Eq. (1) each binary vector $ \mu $ has three components three-qubit system, therefore

\begin{align}
&\mu_{000}=(0,0,0),\mu_{001}=(0,0,1),\mu_{010}=(0,1,0),\nonumber\\
&\mu_{011}=(0,1,1),\mu_{100}=(1,0,0),\mu_{101}=(1,0,1),\nonumber\\
&\mu_{110}=(1,1,0),\mu_{111}=(1,1,1).
\end{align}

Finally, by replacement these values of $ \Gamma $ and $ \mu $ in Eq. (1), the same Eq. (8) graph state is obtained.

\section{\label{sec:level3}GENERALIZED CONCURRENCE}

One of the measurable quantity of entanglement is concurrence. This parameter for the first time is defined by \textrm{Wootters} \textit{et al} for pure and mixed states that have only two qubits \cite{Wootters17}. Then the generalized concurrence of a pure state was defined by \textrm{Mintert} \textit{et al} for $ N $-partite systems as \cite{Carvalho18,Mintert19,Fei20}

\begin{equation}
C_{1,2,\ldots,N}(|\psi\rangle)=2^{1-{\frac{N}{2}}}\sqrt{2^{N}-2-\sum_{\alpha}Tr\rho_{\alpha}^{2}},
\end{equation}

where $ \alpha $ labels as all   different subsystems of the $ N $-partite system and $ \rho_{\alpha} $ are the corresponding reduced density matrices that determined by taking the partial trace of $ \hat{\rho}=|\psi\rangle\langle\psi| $.

By taking partial trace of $ \hat{\rho}=|G\rangle\langle G| $, all of reduced density matrices are calculated. Then by replacing them on the Eq. (11), generalized concurrence is obtained for graph states with three, four and five qubits, which the following relationships stand for them, respectively

\begin{equation}
Tr (\rho_{i}^{2})=Tr (\rho_{jk}^{2}),
\end{equation}

\begin{equation}
Tr (\rho_{ij}^{2})=Tr (\rho_{kl}^{2}), Tr (\rho_{i}^{2})=Tr (\rho_{jkl}^{2}),
\end{equation}

\begin{equation}
 Tr (\rho_{ijk}^{2})=Tr (\rho_{lp}^{2}), Tr (\rho_{i}^{2})=Tr (\rho_{jklp}^{2}),
\end{equation}

where

\begin{equation}
i\neq j\neq k\neq l\neq p \in \{1,2,\ldots,N\}.
\end{equation}

For example, graph state density matrix Eq. (8) is obtained as follows

\begin{align}
&\hat{\rho}=|G_{2}\rangle\langle G_{2}|\nonumber\\
&\,\,\,=\frac{1}{8}\left(
                                      \begin{array}{cccccccc}
                                        1 & 1 & 1 & 1 & 1 & -1 & 1 & -1 \\
                                        1 & 1 & 1 & 1 & 1 & -1 & 1 & -1 \\
                                        1 & 1 & 1 & 1 & 1 & -1 & 1 & -1 \\
                                        1 & 1 & 1 & 1 & 1 & -1 & 1 & -1 \\
                                        1 & 1 & 1 & 1 & 1 & -1 & 1 & -1 \\
                                        -1 & -1 & -1 & -1 & -1 & 1 & -1 & 1 \\
                                        1 & 1 & 1 & 1 & 1 & -1 & 1 & -1 \\
                                        -1 & -1 & -1 & -1 & -1 & 1 & -1 & 1 \\
                                      \end{array}
                                    \right),
\end{align}

then the reduced density matrix $ \rho_{2} $ is calculated as follows

\begin{equation}
\rho_{2}=Tr_{13}(\hat{\rho})=\frac{1}{2}\left(
                         \begin{array}{cc}
                           1 & 1 \\
                           1 & 1 \\
                         \end{array}
                       \right),
\end{equation}

$ \rho_{1} $ and $ \rho_{3} $ are also as follows

\begin{equation}
\rho_{1}=\rho_{3}=\frac{1}{2}\left(
                         \begin{array}{cc}
                           1 & 0 \\
                           0 & 1 \\
                         \end{array}
                       \right),
\end{equation}

as a result, for this state, $ \sum_{\alpha} Tr \rho_{\alpha}^{2}=4 $ and through using Eq. (11), generalized concurrence is as follows

\begin{equation}
C\,(|G_{2}\rangle)=1.
\end{equation}

\section{\label{sec:level4}LOCAL COMPLEMENTATION}

By local complementation of a graph $ G=(V,E) $ at some vertex of $ a\in V $ one obtains an LC-equivalent graph state $ |\tau_{a}(G)\rangle $

\begin{equation}
|\tau_{a}(G)\rangle=U_{a}^{\tau}(G)|G\rangle,
\end{equation}

where

\begin{equation}
U_{a}^{\tau}(G)=\exp(-i\frac{\pi}{4}\sigma_{x}^{a})\prod_{b\in N_{a}}\exp(i\frac{\pi}{4}\sigma_{z}^{b}),
\end{equation}

is a local Clifford unitary and $ N_{a} $ is neighbors of vertex $ a $. Moreover, two graph states $ |G\rangle $ and $ |G^{\prime}\rangle $ are LC-equivalent if and only if the corresponding graphs are related by a sequence of local complementations \cite{Hein09}. To put it simply, $ U_{a}^{\tau}(G) $ in this way acts that an edge between two neighbors of a is deleted if the two neighbors are themselves connected, or an edge is added otherwise.

\section{\label{sec:level5}CLASSIFICATION OF GRAPH STATES UP TO FIVE-QUBIT}

Three, four and five-qubit systems in total, have 1096 different graph states. Some of these states are isomorphic and due to identical amounts of entanglement it is not necessary to calculate the entanglement in all states. By definition isomorphic graphs, it has been found that there are four non-isomorphic graph states with three vertices and there are eleven non-isomorphic graph states with four vertices and there are thirty-four non-isomorphic graph states with five vertices.

For example, the graph state No. 17 in figure 1 contains nine identical graph states that their entanglement is the same \cite{Hein09,Eisert14,Cabello21}. So by calculating the concurrence of non-isomorphic graphs of every system, this parameter can be considered to all graph states of the system.

By using the Mathematica software, all reduced density matrices are obtained for all graph states in figure 1 and then generalized concurrence is calculated. The results of these calculations have been shown in Table I and Table II.
\begin{figure}
  \centering
  \includegraphics[width=15.62 cm]{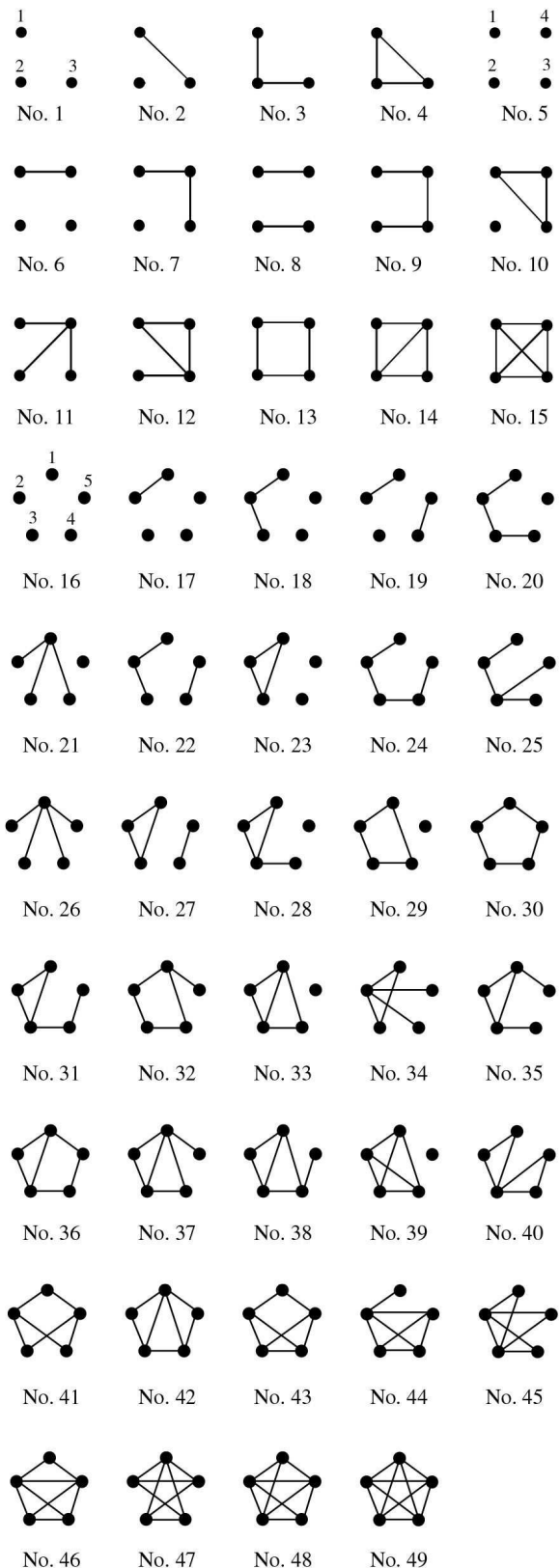}\\
  \caption{The non-isomorphic graphs displayed for three, four and five-qubit graph states. The numbers of qubits are in accordance with the number of graph 1, 5 and 16.}\label{FIG. 1.}
\end{figure}

\begin{table}
  \centering
   \caption{Reduced density matrices.  $(Tr \rho_{i}^{2}=\delta_{i}$,\quad $Tr \rho_{ij}^{2}=\delta_{ij}$,\quad $ i\neq j\in \{1,2,\ldots,N\}$, $\alpha=1$,\quad $\beta=\frac{1}{2}$,\quad $ \gamma= \frac{1}{4}).$  } \label{TABLE I.}
  \begin{tabular}{|l |c| c |c |c |c |c |c| c| c| c| c| c| c |c| c |}
    \hline
\hline
    No. & $\delta_{1}$ & $\delta_{2}$ & $\delta_{3}$ & $\delta_{4}$ & $\delta_{5}$ & $\delta_{12}$ & $\delta_{13}$ & $\delta_{14}$ & $\delta_{15}$ & $\delta_{23}$ & $\delta_{24}$ & $\delta_{25}$ & $\delta_{34}$ & $\delta_{35}$ & $\delta_{45}$ \\

    \hline
    1 &  $\alpha$ & $\alpha$ & $\alpha$ & - & - & - & - & - & - & - & - & - & - & - & - \\
     \hline
    2 &  $\beta$ & $\alpha$ & $\beta$ & - & - & - & - & - & - & - & - & - & - & - & - \\
     \hline
    3 &  $\beta$ & $\beta$ & $\beta$ & - & - & - & - & - & - & - & - & - & - & - & - \\
    \hline
    4 &  $\beta$ & $\beta$ & $\beta$ & - & - & - & - & - & - & - & - & - & - & - & - \\
    \hline
    5 &  $\alpha$ & $\alpha$ & $\alpha$ & $\alpha$ & - & $\alpha$ & $\alpha$ & $\alpha$ & - & - & - & - & - & - & - \\
     \hline
    6 &  $\beta$ & $\alpha$ & $\alpha$ & $\beta$ & - & $\beta$ & $\beta$ & $\alpha$ & - & - & - & - & - & - & - \\
     \hline
    7 &  $\beta$ & $\alpha$ & $\beta$ & $\beta$ & - & $\beta$ & $\beta$ & $\beta$ & - & - & - & - & - & - & - \\
     \hline
    8 &  $\beta$ & $\beta$ & $\beta$ & $\beta$ & - & $\gamma$ & $\gamma$ & $\alpha$ & - & - & - & - & - & - & - \\
     \hline
    9 &  $\beta$ & $\beta$ & $\beta$ & $\beta$ & - & $\gamma$ & $\gamma$ & $\beta$ & - & - & - & - & - & - & - \\
     \hline
    10 &  $\beta$ & $\alpha$ & $\beta$ & $\beta$ & - & $\beta$ & $\beta$ & $\beta$ & - & - & - & - & - & - & - \\
     \hline
    11 &  $\beta$ & $\beta$ & $\beta$ & $\beta$ & - & $\beta$ & $\beta$ & $\beta$ & - & - & - & - & - & - & - \\
     \hline
    12 &  $\beta$ & $\beta$ & $\beta$ & $\beta$ & - & $\gamma$ & $\gamma$ & $\beta$ & - & - & - & - & - & - & - \\
     \hline
    13 &  $\beta$ & $\beta$ & $\beta$ & $\beta$ & - & $\gamma$ & $\beta$ & $\gamma$ & - & - & - & - & - & - & - \\
    \hline
    14 &  $\beta$ & $\beta$ & $\beta$ & $\beta$ & - & $\gamma$ & $\beta$ & $\gamma$ & - & - & - & - & - & - & - \\
    \hline
    15 &  $\beta$ & $\beta$ & $\beta$ & $\beta$ & - & $\beta$ & $\beta$ & $\beta$ & - & - & - & - & - & - & - \\
     \hline
    16 &  $\alpha$ & $\alpha$ & $\alpha$ & $\alpha$ & $\alpha$ & $\alpha$ & $\alpha$ & $\alpha$ & $\alpha$ & $\alpha$ & $\alpha$ & $\alpha$ & $\alpha$ & $\alpha$ & $\alpha$ \\
     \hline
    17 &  $\beta$ & $\beta$ & $\alpha$ & $\alpha$ & $\alpha$ & $\alpha$ & $\beta$ & $\beta$ & $\beta$ & $\beta$ & $\beta$ & $\beta$ & $\alpha$ & $\alpha$ & $\alpha$ \\
     \hline
    18 &  $\beta$ & $\beta$ & $\beta$ & $\alpha$ & $\alpha$ & $\beta$ & $\beta$ & $\beta$ & $\beta$ & $\beta$ & $\beta$ & $\beta$ & $\beta$ & $\beta$ & $\alpha$ \\
     \hline
    19 &  $\beta$ & $\beta$ & $\alpha$ & $\beta$ & $\beta$ & $\alpha$ & $\beta$ & $\gamma$ & $\gamma$ & $\beta$ & $\gamma$ & $\gamma$ & $\beta$ & $\beta$ & $\alpha$ \\
     \hline
    20 &  $\beta$ & $\beta$ & $\beta$ & $\beta$ & $\alpha$ & $\beta$ & $\gamma$ & $\gamma$ & $\beta$ & $\gamma$ & $\gamma$ & $\beta$ & $\beta$ & $\beta$ & $\beta$ \\
     \hline
    21 &  $\beta$ & $\beta$ & $\beta$ & $\beta$ & $\alpha$ & $\beta$ & $\beta$ & $\beta$ & $\beta$ & $\beta$ & $\beta$ & $\beta$ & $\beta$ & $\beta$ & $\beta$ \\
     \hline
    22 &  $\beta$ & $\beta$ & $\beta$ & $\beta$ & $\beta$ & $\beta$ & $\beta$ & $\gamma$ & $\gamma$ & $\beta$ & $\gamma$ & $\gamma$ & $\gamma$ & $\gamma$ & $\alpha$ \\
     \hline
    23 &  $\beta$ & $\beta$ & $\beta$ & $\alpha$ & $\alpha$ & $\beta$ & $\beta$ & $\beta$ & $\beta$ & $\beta$ & $\beta$ & $\beta$ & $\beta$ & $\beta$ & $\alpha$ \\
     \hline
    24 &  $\beta$ & $\beta$ & $\beta$ & $\beta$ & $\beta$ & $\beta$ & $\gamma$ & $\gamma$ & $\gamma$ & $\gamma$ & $\gamma$ & $\gamma$ & $\gamma$ & $\gamma$ & $\beta$ \\
     \hline
    25  & $\beta$ & $\beta$ & $\beta$ & $\beta$ & $\beta$ & $\beta$ & $\gamma$ & $\gamma$ & $\gamma$ & $\gamma$ & $\gamma$ & $\gamma$ & $\beta$ & $\beta$ & $\beta$ \\
     \hline
    26  & $\beta$ & $\beta$ & $\beta$ & $\beta$ & $\beta$ & $\beta$ & $\beta$ & $\beta$ & $\beta$ & $\beta$ & $\beta$ & $\beta$ & $\beta$ & $\beta$ & $\beta$ \\
     \hline
    27 &  $\beta$ & $\beta$ & $\beta$ & $\beta$ & $\beta$ & $\beta$ & $\beta$ & $\gamma$ & $\gamma$ & $\beta$ & $\gamma$ & $\gamma$ & $\gamma$ & $\gamma$ & $\alpha$ \\
     \hline
    28 &  $\beta$ & $\beta$ & $\beta$ & $\beta$ & $\alpha$ & $\beta$ & $\gamma$ & $\gamma$ & $\beta$ & $\gamma$ & $\gamma$ & $\beta$ & $\beta$ & $\beta$ & $\beta$ \\
     \hline
    29 & $\beta$ & $\beta$ & $\beta$ & $\beta$ & $\alpha$ & $\gamma$ & $\beta$ & $\gamma$ & $\beta$ & $\gamma$ & $\beta$ & $\beta$ & $\gamma$ & $\beta$ & $\beta$ \\
     \hline
    30 & $\beta$ & $\beta$ & $\beta$ & $\beta$ & $\beta$ & $\gamma$ & $\gamma$ & $\gamma$ & $\gamma$ & $\gamma$ & $\gamma$ & $\gamma$ & $\gamma$ & $\gamma$ & $\gamma$ \\
     \hline
    31 &  $\beta$ & $\beta$ & $\beta$ & $\beta$ & $\beta$ & $\beta$ & $\gamma$ & $\gamma$ & $\gamma$ & $\gamma$ & $\gamma$ & $\gamma$ & $\gamma$ & $\gamma$ & $\beta$ \\
     \hline
    32 &  $\beta$ & $\beta$ & $\beta$ & $\beta$ & $\beta$ & $\gamma$ & $\gamma$ & $\gamma$ & $\beta$ & $\gamma$ & $\beta$ & $\gamma$ & $\gamma$ & $\gamma$ & $\gamma$ \\
     \hline
    33 & $\beta$ & $\beta$ & $\beta$ & $\beta$ & $\alpha$ & $\gamma$ & $\beta$ & $\gamma$ & $\beta$ & $\gamma$ & $\beta$ & $\beta$ & $\gamma$ & $\beta$ & $\beta$ \\
     \hline
    34 &  $\beta$ & $\beta$ & $\beta$ & $\beta$ & $\beta$ & $\gamma$ & $\beta$ & $\gamma$ & $\gamma$ & $\gamma$ & $\beta$ & $\beta$ & $\gamma$ & $\gamma$ & $\beta$ \\
       \hline
    35 &  $\beta$ & $\beta$ & $\beta$ & $\beta$ & $\beta$ & $\gamma$ & $\gamma$ & $\gamma$ & $\beta$ & $\gamma$ & $\gamma$ & $\gamma$ & $\beta$ & $\gamma$ & $\gamma$ \\
     \hline
    36 & $\beta$ & $\beta$ & $\beta$ & $\beta$ & $\beta$ & $\gamma$ & $\gamma$ & $\gamma$ & $\gamma$ & $\gamma$ & $\gamma$ & $\gamma$ & $\gamma$ & $\gamma$ & $\gamma$ \\
     \hline
    37 &  $\beta$ & $\beta$ & $\beta$ & $\beta$ & $\beta$ & $\gamma$ & $\gamma$ & $\gamma$ & $\beta$ & $\gamma$ & $\beta$ & $\gamma$ & $\gamma$ & $\gamma$ & $\gamma$ \\
     \hline
    38 &  $\beta$ & $\beta$ & $\beta$ & $\beta$ & $\beta$ & $\gamma$ & $\beta$ & $\gamma$ & $\gamma$ & $\gamma$ & $\gamma$ & $\gamma$ & $\gamma$ & $\gamma$ & $\beta$ \\
    \hline
    39 &  $\beta$ & $\beta$ & $\beta$ & $\beta$ & $\alpha$ & $\beta$ & $\beta$ & $\beta$ & $\beta$ & $\beta$ & $\beta$ & $\beta$ & $\beta$ & $\beta$ & $\beta$ \\
    \hline
    40 &   $\beta$ & $\beta$ & $\beta$ & $\beta$ & $\beta$ & $\beta$ & $\gamma$ & $\gamma$ & $\gamma$ & $\gamma$ & $\gamma$ & $\gamma$ & $\gamma$ & $\gamma$ & $\beta$ \\
     \hline
    41 &  $\beta$ & $\beta$ & $\beta$ & $\beta$ & $\beta$ & $\gamma$ & $\beta$ & $\beta$ & $\gamma$ & $\gamma$ & $\gamma$ & $\beta$ & $\beta$ & $\gamma$ & $\gamma$ \\
      \hline
    42 & $\beta$ & $\beta$ & $\beta$ & $\beta$ & $\beta$ & $\gamma$ & $\gamma$ & $\gamma$ & $\gamma$ & $\gamma$ & $\gamma$ & $\gamma$ & $\gamma$ & $\gamma$ & $\gamma$ \\
     \hline
    43 &  $\beta$ & $\beta$ & $\beta$ & $\beta$ & $\beta$ & $\gamma$ & $\gamma$ & $\gamma$ & $\gamma$ & $\gamma$ & $\gamma$ & $\beta$ & $\beta$ & $\gamma$ & $\gamma$ \\
     \hline
    44 &  $\beta$ & $\beta$ & $\beta$ & $\beta$ & $\beta$ & $\gamma$ & $\gamma$ & $\gamma$ & $\beta$ & $\beta$ & $\beta$ & $\gamma$ & $\beta$ & $\gamma$ & $\gamma$ \\
     \hline
    45 &  $\beta$ & $\beta$ & $\beta$ & $\beta$ & $\beta$ & $\gamma$ & $\gamma$ & $\beta$ & $\beta$ & $\beta$ & $\gamma$ & $\gamma$ & $\gamma$ & $\gamma$ & $\beta$ \\
     \hline
    46 &  $\beta$ & $\beta$ & $\beta$ & $\beta$ & $\beta$ & $\gamma$ & $\gamma$ & $\gamma$ & $\gamma$ & $\gamma$ & $\gamma$ & $\beta$ & $\beta$ & $\gamma$ & $\gamma$ \\
     \hline
    47 &  $\beta$ & $\beta$ & $\beta$ & $\beta$ & $\beta$ & $\gamma$ & $\gamma$ & $\gamma$ & $\gamma$ & $\beta$ & $\gamma$ & $\gamma$ & $\gamma$ & $\gamma$ & $\beta$ \\
     \hline
    48 &  $\beta$ & $\beta$ & $\beta$ & $\beta$ & $\beta$ & $\gamma$ & $\gamma$ & $\beta$ & $\gamma$ & $\beta$ & $\gamma$ & $\beta$ & $\gamma$ & $\beta$ & $\gamma$ \\
    \hline
    49  & $\beta$ & $\beta$ & $\beta$ & $\beta$ & $\beta$ & $\beta$ & $\beta$ & $\beta$ & $\beta$ & $\beta$ & $\beta$ & $\beta$ & $\beta$ & $\beta$ & $\beta$ \\
    \hline
  \end{tabular}
\end{table}

\begin{table}
  \centering
  \caption{Classification of graph states up to five-qubit based on generalized concurrence  $(C) $.}\label{TABLE II.}
  \begin{tabular}{l c}
     \hline
 \hline
     No. &\qquad\qquad $C$ \\

     \hline

     1 &\qquad\qquad0 \\
     2 &\qquad\qquad 1 \\
     3,4 &\qquad\qquad 1.2247 \\
     \hline
     5 &\qquad\qquad 0 \\
     6 &\qquad\qquad 1 \\
     7,10 &\qquad\qquad 1.2247 \\
     8,11,15 &\qquad\qquad 1.3229 \\
     9,12,13,14 &\qquad\qquad 1.4142 \\
     \hline
     16 &\qquad\qquad 0 \\
     17 &\qquad\qquad 1 \\
     18,23 &\qquad\qquad 1.2247 \\
     19,21,39 &\qquad\qquad 1.3229 \\
     26,49 &\qquad\qquad 1.3693 \\
     20,28,29,33 &\qquad\qquad 1.4142 \\
     22,27 &\qquad\qquad 1.4577 \\
     25,34,41,44,45,48 &\qquad\qquad 1.5000 \\
     24,31,32,35,37,38,40,43,46,47 &\qquad\qquad 1.5411 \\
     30,36,42 &\qquad\qquad 1.5811 \\
     \hline
      \hline
   \end{tabular}
\end{table}
The presented results in Table II indicate that eight non-isomorphic graph states with three qubits have only three concurrences numeric values and sixty-four non-isomorphic graph states with four qubits have only five concurrences numeric values. 1024 graph states with five qubits have only ten concurrences numeric values as well. Local complementation operator (LC) classifies three, four and five-qubit graph states respectively in three, six and eleven categories \cite{Hein09,Eisert14}.
Comparing the classifications trough two methods of generalized concurrence and LC operator this result is obtained that the three-qubit system number of categories graph states are identical with both methods, but in the four and five-qubit system of classes obtained through generalized concurrence is less than the LC-rule. Moreover, it cannot be found any two states in the classification performed by the LC operator placed in a category which has two different values generalized concurrences.
For example, according to classification by LC-rule, graph states No. 11 and No. 15 are in a category and has also equal generalized concurrence values. On the other hand, despite graph No. 8 has a different category LC-rule with graphs No. 11 and No. 15, the value of the generalized concurrence is identical with graphs No. 11 and No. 15. As a result of the generalized concurrence, quantum graph states are relatively well differentiated and categorized, but it was not able to distinguish between all graph states.
In other words, this classification that has been carried out with the generalized concurrence is not in contradiction with the LC-rule. So because there is no contradiction, as well as relatively high resolution, generalized concurrence is considered a reliable quantitative for measuring entanglement.

\section{\label{sec:level6}CONCLUSIONS}

Using the definitions and concepts of the mathematical graph and replacing each vertex as a qubit and considering each edge as an interaction between two qubits, we obtained all graph states up to five-qubit. Then the entanglement in each of these states has been calculated by generalized concurrence. We offer the new classification of entangled states up to five qubits under measuring by this measurement.
Using the method of comparing the results of this classification with equivalence classes of these graph states under local complementation (LC) operator, the results show that the new classification by the generalized concurrence is not in contradiction with the classification under LC-rule. Thus generalized concurrence is considered as a reliable quantitative for measuring entanglement. The proposed approach can be used to recognize the proper performance of each new quantity suggested for measuring entanglement.

\bibliography{Article}

\end{document}